\documentclass{elsart1p}
\def\pt{${\rm p_{_{T}}}$}
\def\ptt{$p_T^t$}

\def\snn{${\rm \sqrt{s_{_{NN}}}}$ =}
\def\p{$p\mathord{+}p$}
\def\pta{$p_T^{a}$}
\def\po{$\pi^{0}$}
\def\Au{Au$\mathord{+}$Au}
\def\Cu{Cu$\mathord{+}$Cu}
\def\dAu{d$\mathord{+}$Au}
\def\pt{${\rm p_{_{T}}}$}
\def\raa{$R_{AA}$}
\def\rda{$R_{AA}$}
\def\iaa{$I_{\rm AA}$}
\def\Np{$N_{\rm part}$}

\usepackage{graphicx}
\usepackage{amssymb}
\def\FigureOne{
\begin{figure*}
\begin{center}
\resizebox{1.0\textwidth}{!}{\includegraphics{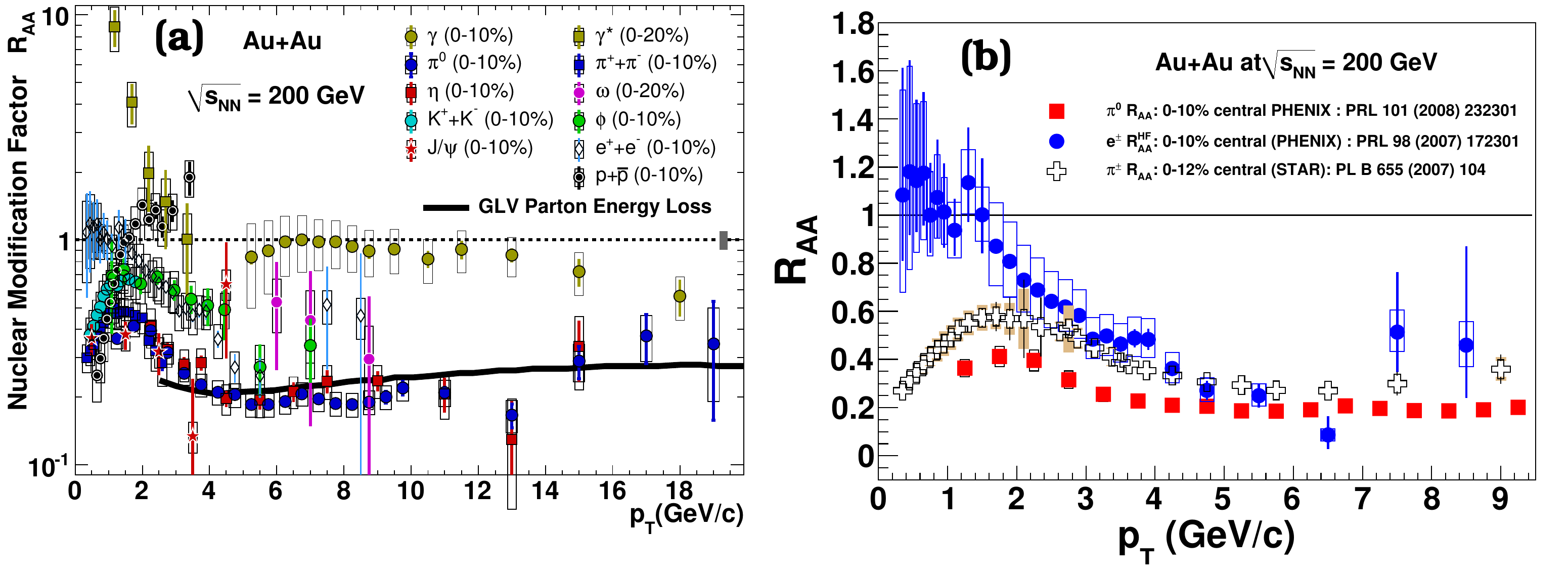}}
\caption[]{Compilation of data of nuclear modification factor ${\rm
    R_{AA}}$ from \Au\ collisions at \snn\ 200 GeV
  \cite{CompilationRaa}. Panel (a) nuclear modification factor,
  $R_{AA}$, for identified particles measured by PHENIX, given as a
  function of transverse momentum, $p_{T}$. The continous curve
  corresponds to the GLV theory~\cite{Vitev}. Panel (b) represents the
  nuclear modification factor of heavy-flavor electrons ${\rm
    R_{AA}^{HF}}$ compared with the ${\rm R_{AA}}$ of $\pi^{0}$ and
  $\pi^{\pm}$.}
\label{fig1}
\end{center}
\end{figure*}
}
\def\FigureTwo{
\begin{figure*}
\begin{center}
\resizebox{0.8\textwidth}{!}{\includegraphics{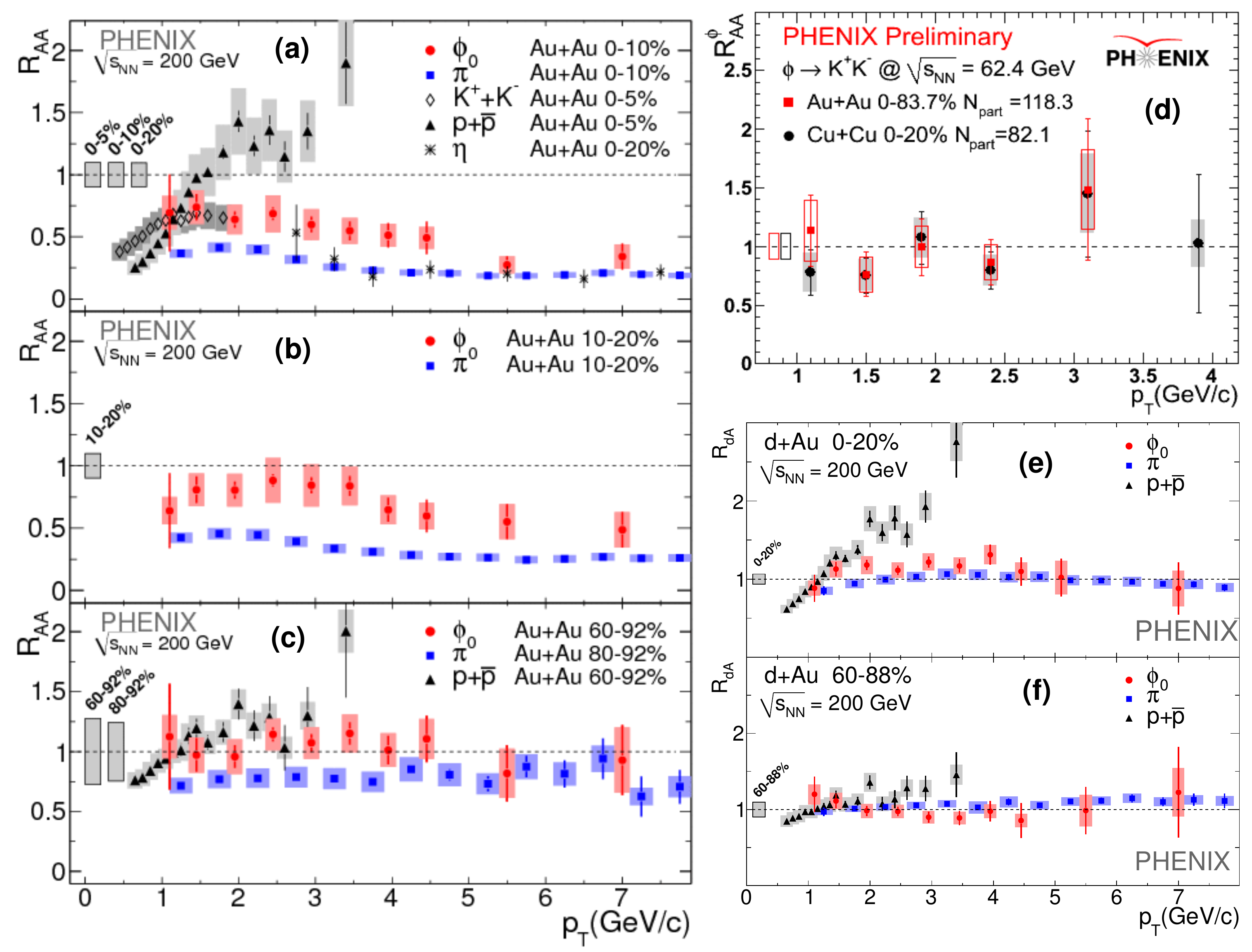}}
\caption[]{Panels (a), (b), and (c) show the \raa\ for different
  particle species obtained from \Au\ collisions at \snn\ 200
  GeV. Panel (d) shows the $\phi$ \raa\ for centrality bin selected
  such as the \Np\ is similar in \Au\ and \Cu\ collisions at \snn\
  62.4 GeV. Panels (e) and (f) show the \raa\ for different particle
  species, including the $\phi$'s obtained from \dAu\ collisions at
  \snn\ 200 GeV.}
\label{fig2}
\end{center}
\end{figure*}
}
\def\FigureThree{
\begin{figure*}
\begin{center}
\resizebox{0.7\textwidth}{!}{\includegraphics{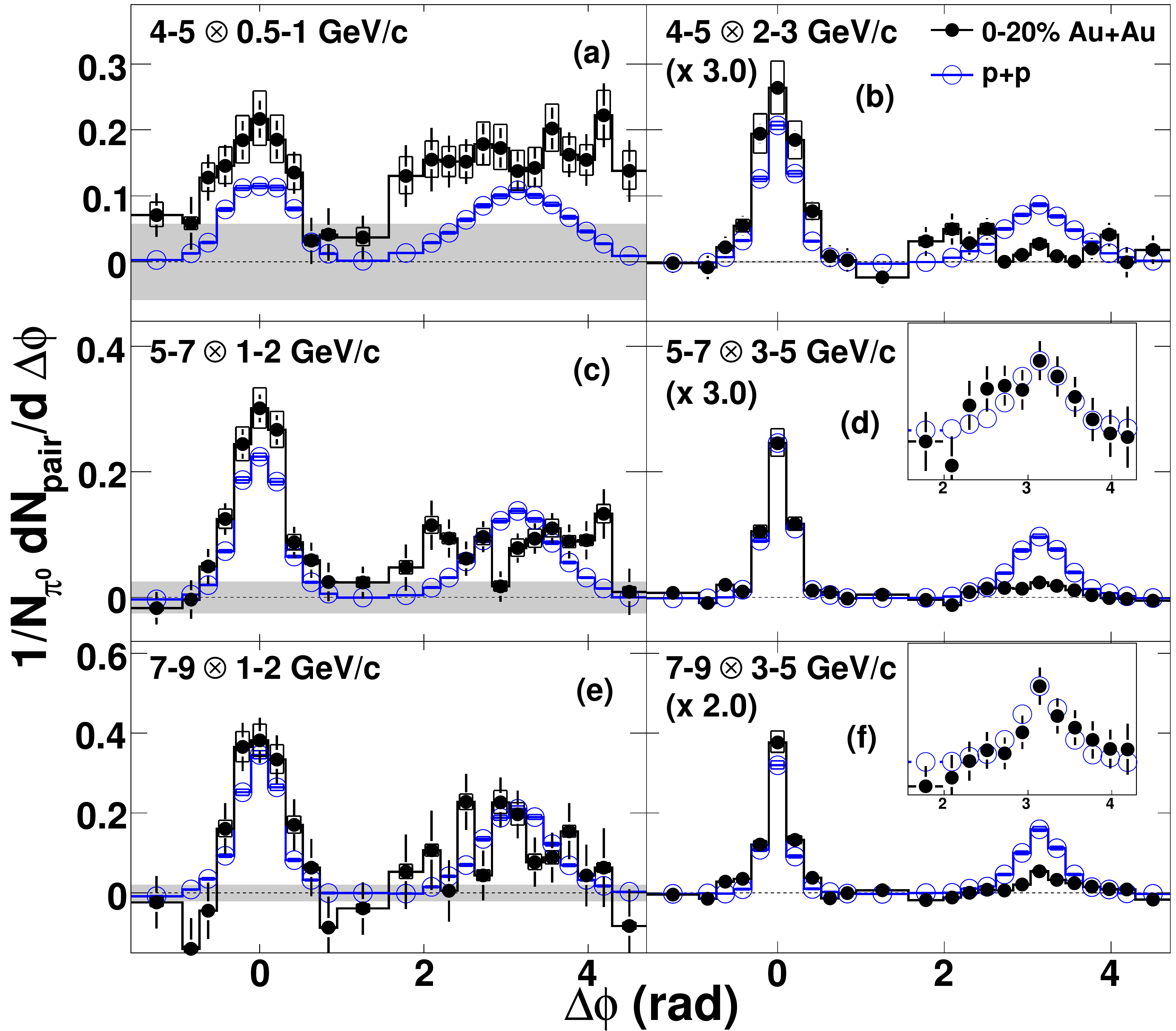}}
\caption[]{Per-trigger jet pair yield versus $\Delta\phi$ for selected
  $\pi^{0}$ trigger and $h^\pm$ partner $p_T$ combinations ($p_T^t$
  $\times$ $p_T^{a}$ in \Au\ (solid symbols) and \p\ (open symbols)
  collisions. The depicted \Au\ systematic uncertainties include
  point-to-point correlated background level and modulation
  uncertainties (shown respectively as gray bands and open boxes). For
  comparing shape, the insets show the away-side distributions scaled
  to match at $\Delta\phi = \pi$.}
\label{fig3}
\end{center}
\end{figure*}
}
\def\FigureFour{
\begin{figure*}
\begin{center}
\resizebox{0.6\textwidth}{!}{\includegraphics{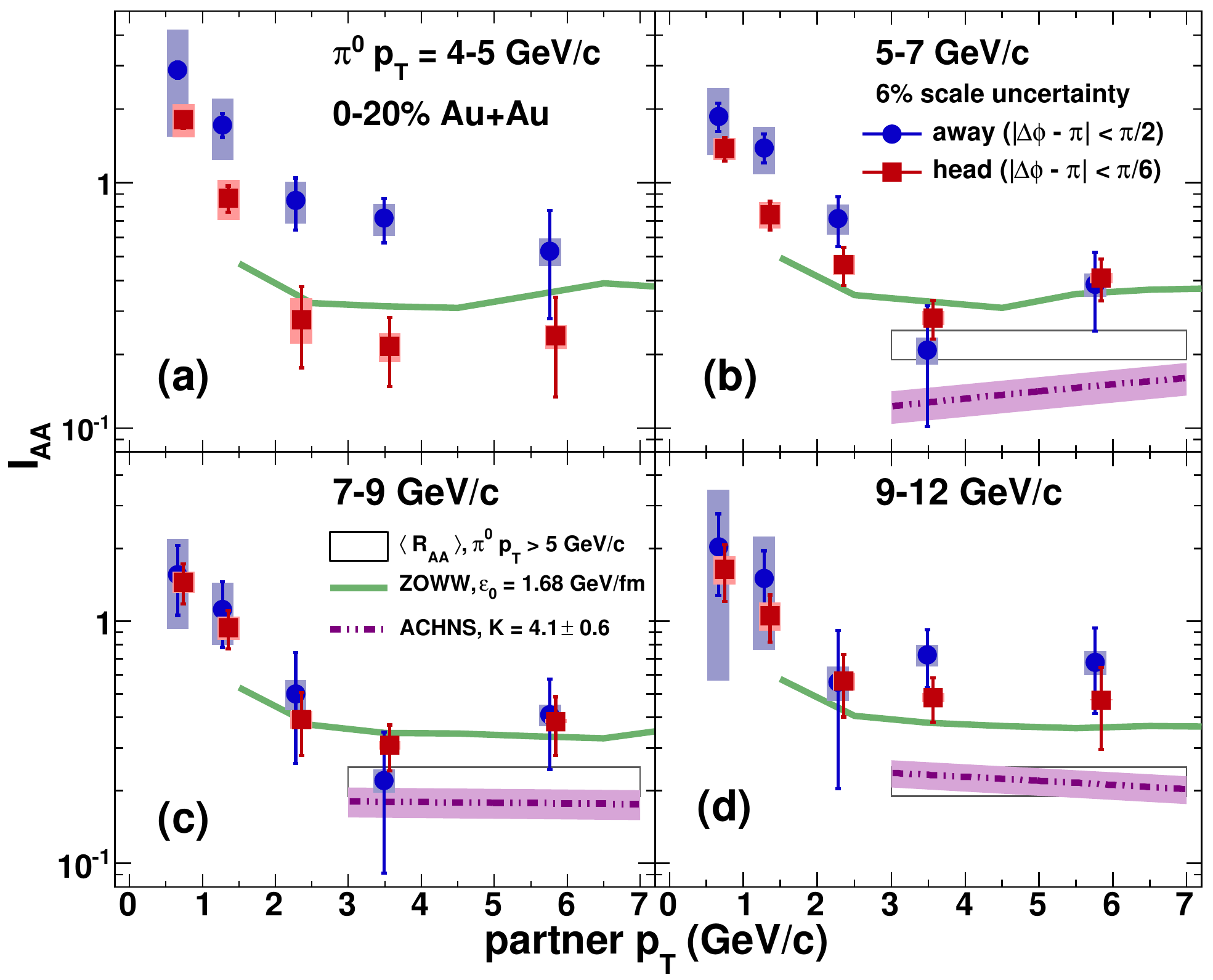}}
\caption[]{Away-side $I_{\rm AA}$ for a narrow ``head''
  $|\Delta\phi-\pi|<\pi/6$ selection (solid squares) and the entire
  away-side, $|\Delta\phi-\pi|<\pi/2$ (solid circles) versus $h^\pm$
  partner momentum for various \p\ trigger momenta. Calculations are shown from
  two different predictions for the head region in
  applicable $p_T$ ranges~\cite{correlations}. A point-to-point uncorrelated 6\%
  normalization uncertainty (mainly due to efficiency corrections)
  applies to all measurements.}
\label{fig4}
\end{center}
\end{figure*}
}
\def\FigureFive{
\begin{figure*}
\begin{center}
\resizebox{1.0\textwidth}{!}{\includegraphics{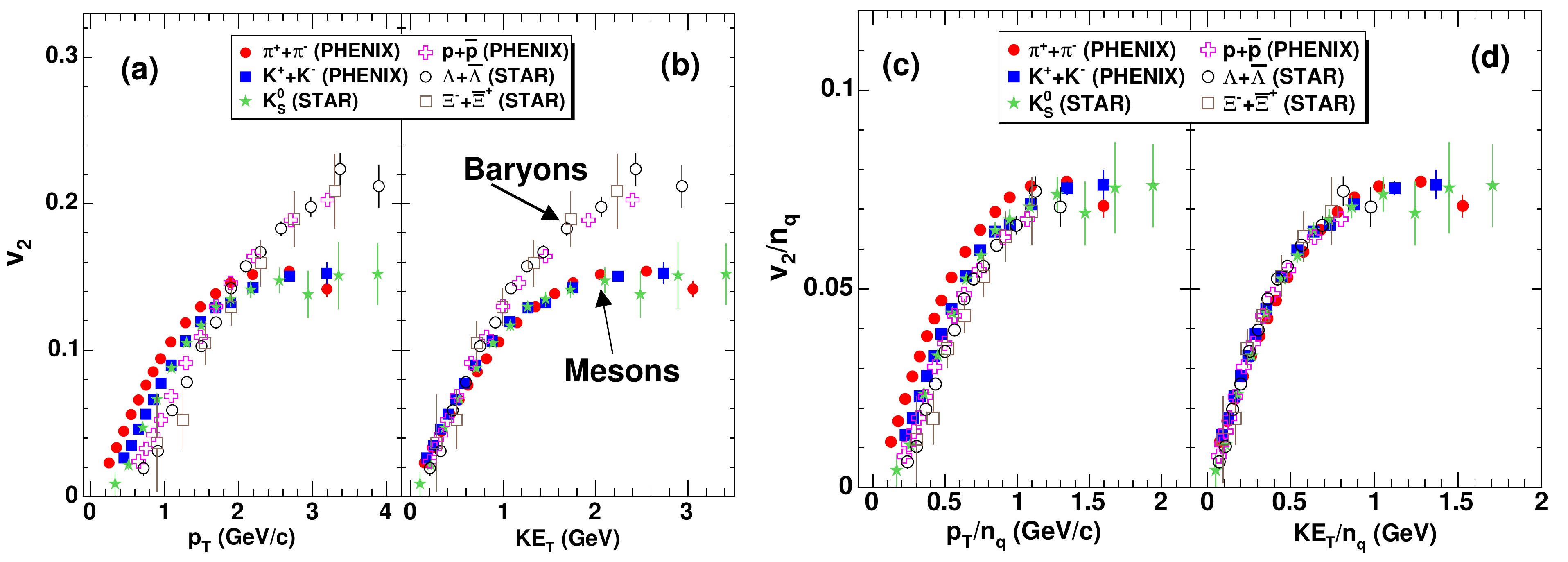}}
\caption{Identified hadron anisotropy: Panels (a) as a function of
  transverse momentum \pt, panel (b) as a function of kinetic energy,
  panel c) as a function of scaled \pt/${\rm n_{q}}$ and, panel d) as
  a function of scaled transverse kinetic energy (${\rm m_{T}}$ -
  mass)/${\rm n_{q}}$.  n$_{q}$ is the number of quarks valence in a
  given hadron (for mesons, ${\rm n_{q}}$ $=$ 2; and, for baryons:
  ${\rm n_{q}}$ $=$ 3). All data are from minimum-biased \Au\
  collisions at \snn\ 200 GeV.}
\label{fig5}
\end{center}
\end{figure*}
}
\def\FigureSix{
\begin{figure*}
\begin{center}
\resizebox{0.5\textwidth}{!}{\includegraphics{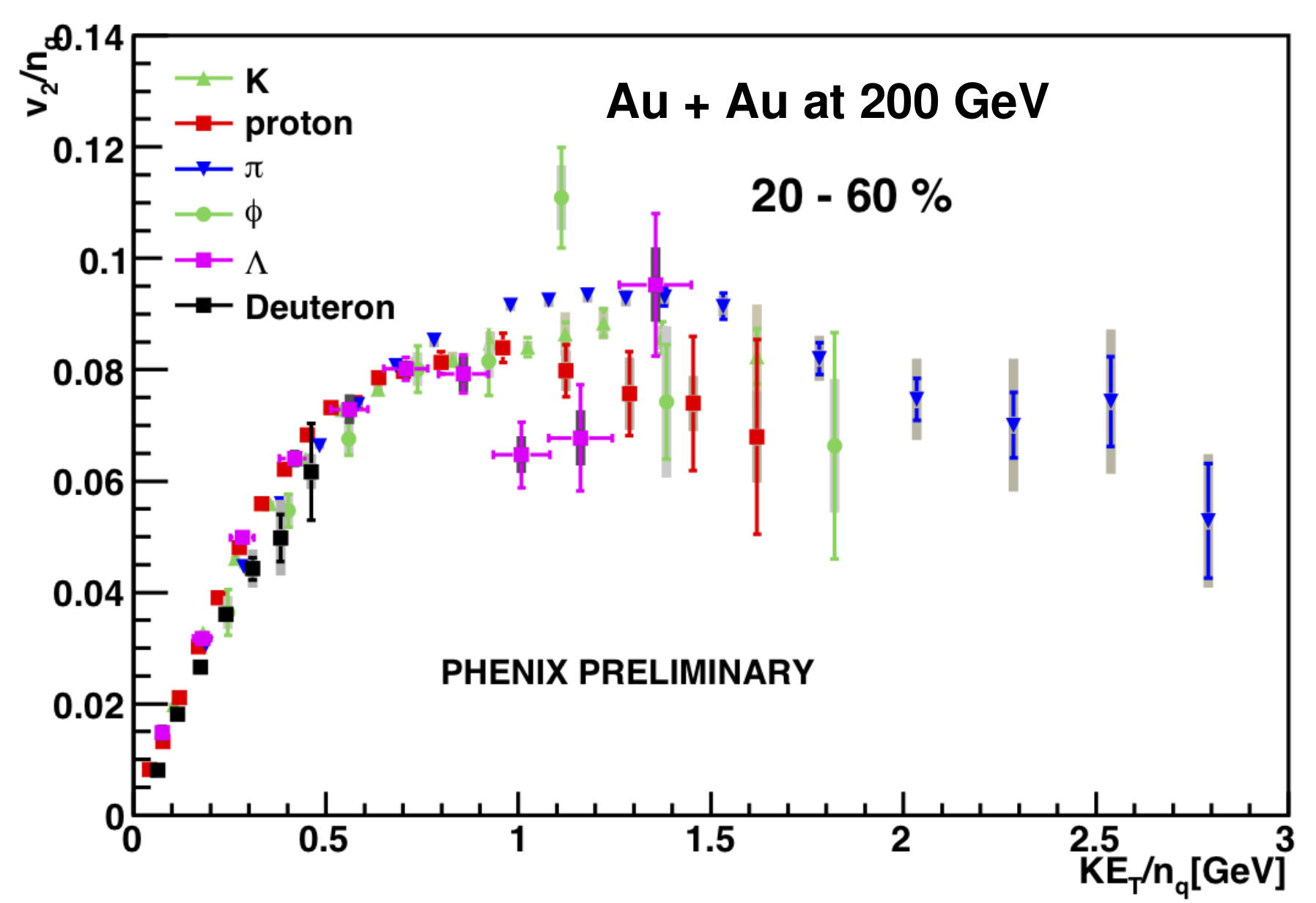}}
\caption{panel (a) shows scaled $v_2$/${\rm n_{q}}$ dependence of the
  transverse kinetic energy ${\rm KE_{T}}$/${\rm n_{q}}$ for $\phi$
  mesons compared with different particle species obtained from \Au\
  at \snn\ 200 GeV (for mesons, ${\rm n_{q}}$ $=$ 2 ; and, for
  baryons: ${\rm n_{q}}$ $=$ 3).}
\label{fig6}
\end{center}

\end{figure*}
}
\def\FigureSeven{
\begin{figure*}
\begin{center}
\resizebox{0.8\textwidth}{!}{\includegraphics{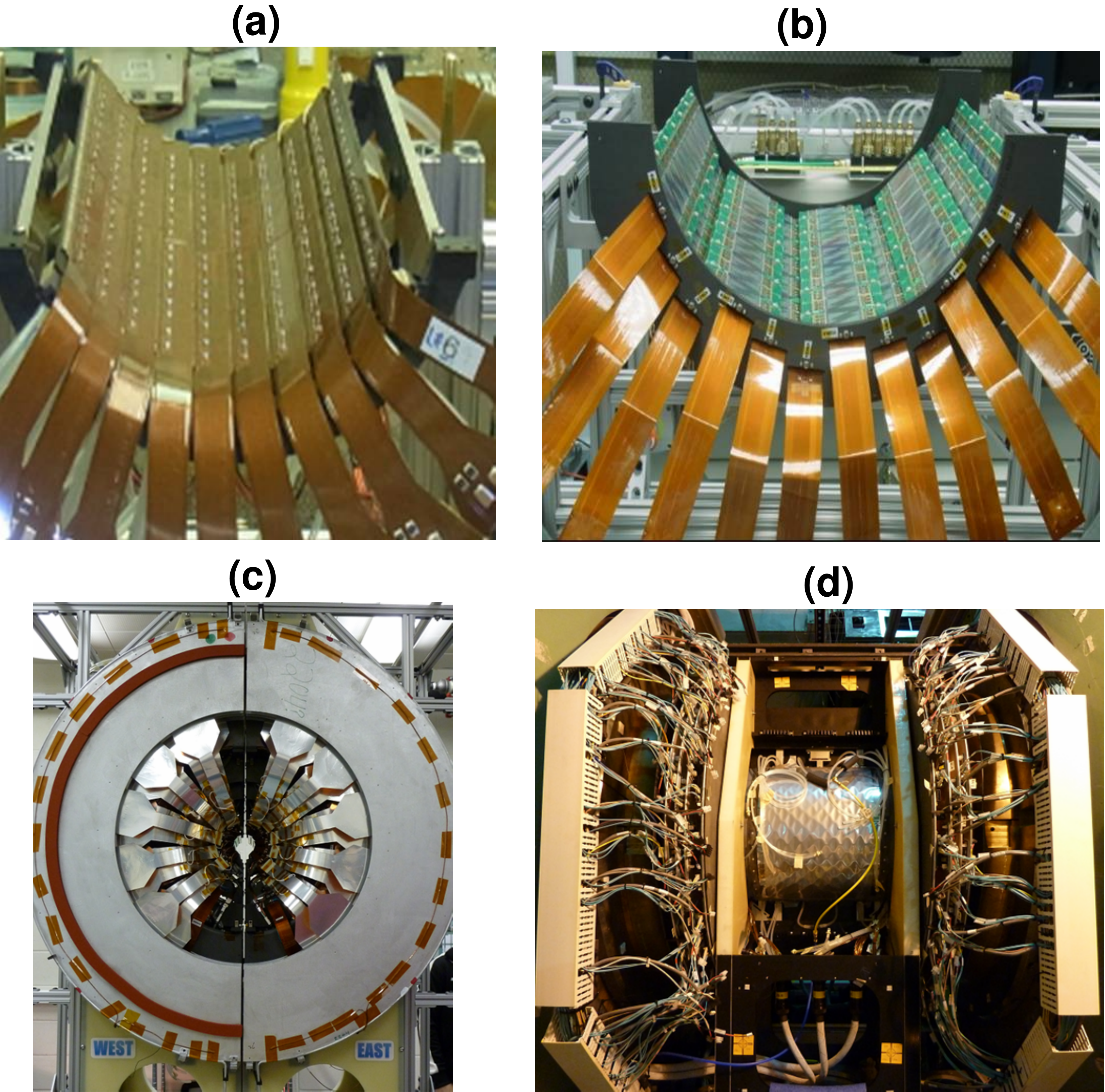}}
\caption{The PHENIX Silicon Vertex Tracker. Panel (a) and (b) show the
  ladders of half barrels of pixel and stripixel detectors,
  respectively. Panel (c) shows VTX detector including the readout,
  assembled and surveyed at the laboratory. Panel (d) shows VTX shows
  the VTX installed around the PHENIX's interaction point and cabled
  to the readout, in December 2010.}
\label{fig7}
\end{center}
\end{figure*}
}
\begin{document}
\begin{frontmatter}
\title{Recent Results from PHENIX Experiment at RHIC: Exploring the QCD Medium}
\author{Rachid Nouicer, for the PHENIX Collaboration}
\address{Physics Department, Brookhaven National Laboratory, Upton, NY 11973-5000, USA}
\vspace*{-0.5cm}
\begin{abstract}
  We review some important results from the PHENIX experiment at
  RHIC. They were obtained in a unique environment for studying QCD
  bulk matter at temperatures and densities that surpass the limits
  where hadrons exist as individual entities, so raising to prominence
  the quark-gluon degrees of freedom. We present measurements of
  nuclear modification factors for neutral pions, light favors
  (strange hadrons), direct-photons and non-photonic electrons from
  decays of particles carrying charm or beauty quarks. We interpret
  the large suppression of hadron production at high transverse
  momenta as resulting from a large energy loss by the precursor parton
  on its path through the dense matter, primarily driven by gluon
  radiation. This dense QCD matter responds to energy loss in a
  pattern consistent with that expected from a hydrodynamic
  fluid. Further, its elliptic flow measurements suggest that the
  hadronization of bulk partonic matter exhibits collectivity with
  effective partonic degrees of freedom. The results are shown as a
  function of transverse momentum, centrality in different collision
  systems and energies.
\end{abstract}
\end{frontmatter}
\vspace*{-0.3cm}
\section{Physics motivation and RHIC achievements}
\vspace*{-0.3cm}
Quantum Chromodynamics (QCD) is considered the fundamental theory for
 strong interactions. According to it, hadronic matter under extreme
dense, hot conditions must undergo a phase
transition~\cite{Shuryak80} to form a Quark Gluon Plasma (QGP) in
which quarks and gluons no longer are confined to the size of a
hadron. Recent results from lattice QCD at finite
temperature~\cite{Satz2003,Karsh2002} reveal a rapid increase in the
number of degrees of freedom associated with this deconfinement of
quarks and gluons from their hadronic chains. The transition point is at
a temperature ${\rm T \approx 170\ MeV}$ and energy density of ${\rm
  \epsilon \approx 1\ GeV/fm^{3}}$. Under the same conditions, chiral
symmetry is restored~\cite{Satz2003}. Therefore, our experiments search
for signatures of both QGP formation, and the in-medium effects of
the hadrons' properties. Purportedly, these required high densities
could be achieved via relativistic heavy ion
collisions~\cite{Greiner1975}.

Under the RHIC (Relativistic Heavy Ion Collider) project, an
accelerator was constructed at Brookhaven National Laboratory (BNL)
from 1991 to 1999. RHIC was designed as a heavy-ion machine, that
would support the collision of a wide range of nuclei over a large
range of energies. Already, gold--gold, copper-copper and
deuteron--gold collisions were attained at energies from \snn\ 7.7 to
200 GeV. Further, a polarized capability was added to RHIC allowing
transverse and longitudinal polarized protons to collide at energies
from \snn\ 200 to 500 GeV. Earlier, RHIC researchers made a major
physics discovery, namely the creation, in high-energy central
gold-gold collisions, of a new form of matter, dense and strongly
interacting, called the strongly coupled quark-gluon plasma, or sQGP.
This finding was rated the top physics story of 2005; the four
experiments at RHIC, BRAHMS, PHENIX, PHOBOS, and STAR, published,
evidence of the existence of this new form of matter \cite{RHIC, RHICNouicer}. The RHIC accelerator also met and surpassed its
specifications; namely, it attained its energy goals, and exceeded, by
a factor of 2, its heavy-ion luminosity goals, and its polarized
proton luminosity by a factor of 5.

In this article, I highlight some of recent PHENIX experiment results
underlying these major experimental observations. I start by
discussing our hard-probe measurements; high-\pt hadron suppression,
including heavey-flavors and di-jet fragment azimuthal correlations
probes. These measurements of hard probes afford direct signatures of
highly interacting dense matter created at the RHIC.  I then present our
measurements of elliptic flow that are an indirect signature of the
existence of partonic matter. These measurements, for different particle
species, are given as a function of collision centrality, energy, and
system sizes.
\FigureOne%
\vspace*{-0.3cm}
\section{High-\pt hadron suppression :  Jet quenching}
\vspace*{-0.3cm} In heavy ion collisions (from AGS to RHIC energies)
hadron production at the mid-rapidity region (${\rm |y| <}$ 1.5) rises
with increasing collision energy. At RHIC, the central zone is almost
baryon free~\cite{BrahmsBaryon}. The large particle production is
dominated by pair production, and the energy density seems to exceed
significantly that required for QGP formation~\cite{RHIC}. The PHENIX
experiment revealed the suppression of the high transverse momentum
component, \pt\ of hadron spectra at the mid-rapidity region in
central Au+Au collisions compared to the scaled momentum spectra from \p\
collisions at the same energy, \snn\ 200 GeV \cite{RHIC}. This effect,
originally proposed by Bjorken, Gyulassy and others \cite{Bjorken}
rests on the expectation of a large energy loss from high momentum
partons scattered during the initial stages of collisions in a medium with
a high density of free color charges. According to QCD theory, colored
objects may lose energy by the Bremsstrahlung radiation of gluons
\cite{Gaard}. Such a mechanism would strongly degrade the energy of
leading partons, as reflected in the reduced transverse momentum of
leading particles in the jets emerging after their fragmentation into
hadrons. Figure~\ref{fig1} compilated the data
\cite{CompilationRaa} for the nuclear modification factors measured in
\Au\ collisions at \snn\ 200 GeV. The nuclear modification factor is
defined as:
\begin{equation}
{\rm  R_{AA} (p_{T})= \frac{d^{2}N^{^{A+A}}/dp_{T}d\eta}{N_{bin}\
    d^{2}N^{^{p+p}}/dp_{T}d\eta}}
\end{equation}
It involves scaling the measured distributions of nucleon-nucleon
transverse momentum by the number of expected incoherent binary
collisions, ${N_{bin}}$ \cite{RAAdefinition}. In the absence of any
modifications caused by the `embedding' of elementary collisions in a
nuclear collision, we expect ${\rm R_{AA}}$ = 1 at high-\pt. At low
\pt, where particle production follows a scaling with the number of
participants, the above definition of ${\rm R_{AA}}$ leads to ${\rm
  R_{AA}}$ $<$ 1 for \pt $<$ 2 GeV/c. Figure~\ref{fig1}(a) summarizes
the present status of ${\rm R_{AA}}$ for neutral pions, light favors
(strange hadrons), direct-photons and non-photonic electrons from
heavy quarks decays in central \Au\ collisions at \snn\ 200 GeV.  The
${\rm R_{AA}}$ for direct photons are not suppressed because they do
not interact strongly with the medium. The ${\rm R_{AA}}$ for both
${\rm \pi^{0}}$ and ${\rm \eta}$'s mesons exhibit the same suppression
relative, by a factor of 5, to the point-like scaled \p\ data appear
to be constant for \pt $>$ 4 GeV/c, while the ${\rm \eta}$ mass is
much larger than that of ${\rm \pi^{0}}$. This observation points to
partonic nature of suppression. These data of ${\rm R_{AA}}$ were
described by theoretical calculations of the partons' energy loss in the
matter created in Au+Au collisions \cite{Vitev1}. From these
theoretical frameworks, we learned that the gluon density ${\rm
  dN_{g}}$/dy must be approximately 1000, and the energy density of
the matter created in the most central collisions must be
approximately 15 GeV/fm$^{3}$ to account for the large suppression
observed in the data \cite{Vitev2,Vitev3}.

PHENIX measured heavy flavor production via semi-leptonic decays, and
identified electrons from the decays of ${D}$- and ${B-}$mesons. In the
mid-rapidity region, electron identification largely is based on using
the Ring Imaging Chernkov detector (RICH), in conjunction with a
highly granular electromagnetic calorimeter (EMC). Their momentum is
derived from the curvature of the track (due to a magnetic field up to
1.15 Tesla) reconstructed from the drift and pad-chambers. A major
difficulty in electron analyzes is that there are many sources of
electrons, other than the semi-leptonic decays of heavy flavor mesons.
Figure~\ref{fig1}(b) compares the nuclear modification factor R$_{AA}$
of heavy flavor electrons to $\pi^{0}$ data, and to the $\pi^{\pm}$ data
obtained from central \Au\ collisions at \snn\ 200 GeV
\cite{CompilationRaa,HeavyPHENIX,HeavySTAR}. We observe clear
suppression of heavy flavor electrons in central events in
high-\pt. For \pt $>$ 4 GeV/c, the R$_{AA}$ of heavy flavors is
surprisingly similar to that for $\pi^{0}$ and $\pi^{\pm}$.
\FigureTwo%

An interesting result of the nuclear modification factor in heavy ion
collisions concerns the $\phi$ meson, whose mass is close to that
of a proton and ${\rm \Lambda}$. The $\phi$ meson does not
participate as strongly as others do in hadronic interactions, nor are
$\phi$ mesons formed via the coalescence-like $K^{+} + K^{-}$
process in high energy collisions \cite{StarKprocess}. 
Figure~\ref{fig2} (a), (b), and (c) show the \raa\ for $\phi$,
\po, proton, kaon and $\eta$, all measured in \Au\
collisions at \snn\ 200 GeV ~\cite{PHENIXphi}. The $\phi$ mesons exhbit a different
suppression pattern than the lighter non-strange mesons and
baryons. For central collisions (Figure~\ref{fig2} (a)) the $\phi$'s
\raa has less suppression that \po\ and $\eta$ in the intermediate \pt
range of 2 $<$ \pt (GeV/c) $<$ 5. At higher \pt\ (\pt\ $>$ 5 GeV/c),
the $\phi$'s \raa\ approaches and becomes comparable to the \po\ and
$\eta$s'\ \raa. These futures remain true for all centralities up to
the most peripheral collisions, as displayed in Figure~\ref{fig2}
(c). This panel shows that the \po\ is slightly suppressed in
peripheral \Au\ collisions whereas the $\phi$ is not suppressed. The
kaon data cover only a very limited range at low \pt; nevertheless, in
this range they seem to follow the \raa\ trend of the $\phi$ better
than that of \po\ and $\eta$ for central \Au\ collisions. The
comparison with baryons (protons and anti-protons) shows different
pattern. For central collisions, the protons show no suppression but rather an
enhancement at \pt\ $<$ 1.5 GeV/c, whereas the $\phi$ mesons are
suppressed. We observed the similarity of the \raa\ of protons and $\phi$
mesons for most peripheral collisions.

Figure~\ref{fig2}(d) compares the \raa\ of $\phi$ in \Au- and \Cu-
collision in two centrality bins corresponding approximately to the
same number of participants in the two systems. Under this condition,
there is no difference in the \raa\ of $\phi$ between the two systems,
indicating that the level of the suppression, when 
averaged over the azimuthal angle, scales with the average size of the nuclear overlap,
regardless of the details of its shape. Cold nuclear matter effects
also contribute to the differences in the hadron suppression factor in
\Au\ collisions. Figures~\ref{fig2}(e) and \ref{fig2}(f) compare the
\rda\ for $\phi$ and \po\ and proton in central and peripheral \dAu\
collisions. For both centralities, the \rda\ for $\phi$ and \po\ are
similar, indicating that cold nuclear matter effects are not
responsible for the differences between them that are evident in \Au\
collisions.  \FigureThree \FigureFour \vspace*{-0.3cm}
\section{Di-jet fragment azimuthal correlations}
\vspace*{-0.3cm}
The study of the production of high transverse momentum hadrons in heavy ion
collisions at RHIC provides an experimental probe of the QCD matter in
the densest stage of the collisions \cite{Jacobs2005}. In particular,
two-hadron azimuthal correlations support the assessment of
back-to-back, hard-scattered partons that propagate in the medium
before fragmenting into jets of hadrons, thereby serving as a
tomographic probe of the medium. It already was observed in
di-hadron correlations from central Au+Au collisions that both the
shape of the relative azimuthal angular distribution, and the yield of
jet-like fragment pairs can depart significantly from those of
$p\mathord{+}p$ collisions~\cite{2005ph,ppg083}. The underlying
mechanisms for jet modification are not fully understood, but
partonic energy loss by QCD radiative processes and collisions with
medium constituents, as well as the evolution of the lost energy,
should contribute to modifying the single and pair yields of
hadrons associated with the jets.

Figure~\ref{fig3} shows the resulting per-trigger jet pair yields for
selected trigger-partner combinations in \p, and the 20\%
most central \Au\ collisions~\cite{correlations}.  On the near side, the
widths in central \Au\ are comparable to \p\
over the full \ptt\ and \pta\ ranges, while the yields are slightly
enhanced at low \pt, matching \p\ as \pt\ increases, while on
the opposite, qualitatively we observe that for low \ptt\ and
low \pta\ the Au$\mathord{+}$Au jet peaks are strongly broadened and
non-Gaussian.  In contrast, at high \ptt\ and high \pta\ the yield 
substantially is suppressed, but the shape appears consistent with the
measurement, as in the $p\mathord{+}p$ case. To study the shape
modification measurement, we determined the away-side integrated yield. This
modification of the determined jet yield in central collisions,
shown in Figure~\ref{fig4}, is measured by $I_{\rm AA}$ such that:
 \begin{equation}
I_{\rm AA} =
\left( \int{\frac{dN^{\rm pair}}{N^{t}}} \right)_{\rm A+A} \left/
\left( \int{\frac{dN^{\rm pair}}{N^{t}}} \right)_{p+p}
\right.
\end{equation}
Away-side $I_{\rm AA}$ values for \ptt\ $>$ 7 GeV/$c$ tend to fall
with \pta\ for both the full away-side region ($|\Delta\phi-\pi| <
\pi/2$) and for a narrower ``head'' selection ($|\Delta\phi-\pi| <
\pi/6$) until \pta\ $\approx$ 2--3 GeV/c, above which they become
roughly constant. The yield enhancement at \ptt\ $>$ 7 GeV/$c$ and
\pta $<$ 2 GeV/$c$ is modest and occurs without significant
modification of the shape. When \ptt\ decreases, the away-side $I_{\rm
  AA}$ of the two angular selections differs as their shape
changes. Figure~\ref{fig4} also shows the \po\ \raa\ for \pt\ $>$ 5
GeV/$c$~\cite{Adare:2008qa}. The comparison reveals that \iaa\ is
consistently higher than \raa. This feature probably reflects a
few competing effects. Selecting high-\pt\ trigger \p\ is
expected to bias hard scattering towards the medium surface.
Thus, away-side partons have a long average path length through the
medium and consequently lose more energy.  However, this feature does not
require that the \iaa\ be lower than \raa. The away-side conditional
spectrum falls less steeply than the inclusive hadron spectrum and so
the same spectral shift will more strongly reduce \raa. More details
on Figure~\ref{fig4} are given in Ref.~\cite{correlations}.
\vspace*{-0.3cm}
\section{Hadronization of bulk partonic matter}
\vspace*{-0.3cm} The anisotropic flow of hadrons was studied
extensively in nucleus-nucleus collisions at the SPS and RHIC as a
function of pseudorapidity, centrality, transverse momentum, and
energy \cite{RHIC,FlowSPS,RachidQM2006}. In non-central collisions of
heavy ions at high energy, the configuration of space anisotropy is
converted into a momentum space anisotropy. The dynamics of the
collision determine the degree of this transformation. For a symmetric
system, like \Au, the second Fourier expansion is a good
parameterization of anisotropy. At RHIC, a strong anisotropic flow
(v$_{2}$) was observed for all hadrons measured. Comparing the
data obtained at the mid-rapidity region presented by the STAR, PHOBOS,
and PHENIX collaborations \cite{RHIC,RachidQM2006} to the hydrodynamic
model \cite{hydro} affording strong evidence that the originated medium
behaves as a near ideal fluid.  
\FigureFive%
\FigureSix%

Figure~\ref{fig5}(a) and (b) show the anisotropic flow distributions,
v$_{2}$, for identified hadrons obtained in minimum-bias \Au\
collisions at \snn\ 200 GeV~\cite{PhenixFlow}. The values for neutral Kaons
($K{^{0}}{_{s}}$), lambdas ($\Lambda$), and the cascade ($\Xi$) are
taken from Ref.\cite{StarFlow}. In the lower \pt\ region, \pt\ $\le$ 2
GeV/c, the value of v$_{2}$ is inversely related to the mass of the
hadron, that is, it is characteristic of hydrodynamic collective
motion in operation.  At the intermediate \pt\ region, the dependence
is different. Instead of a mass dependence, there seems to be a hadron
type dependence, clear splitting into a meson branch and a baryon
branch. This observation is well demonstrated in Figure.~\ref{fig5}(b)
when v$_{2}$ is shown as a function of the transverse kinetic
energy. To include the effect of collective motion, the \pt\ was
transfered to the transverse kinetic energy ${\rm KE_{T}}$ $\equiv$
${\rm m_{T}-}$mass $=$ ${\rm \sqrt{p^{2}_{T}-m^{2}}-m}$ where m is the
mass of the particle.  To demonstrate the scaling properties of
v$_{2}$, the following transformation was performed in which we scaled
the measured v$_{2}$ by the number of valence quarks in a given
hadron. For mesons and baryons, respectively, they are ${\rm n_{q}}$
$=$ 2 and 3. The \pt and ${\rm KE_{T}}$ also were scaled with the same
${\rm n_{q}}$; Figures~\ref{fig5}(c) and \ref{fig5}(d), respectively,
depict the outcome.  All of the hadron v$_{2}$ scaled nicely up to
(${\rm m_{T}}$ $-$ mass)/${\rm n_{q}}$ $\sim$ 1.2. At high-\pt, the
values of v$_{2}$ appear to fall.  These observations about scaling
reveal that before hadronization, quarks already have acquired the
collective motion v$_{2}$, and that when they coalesce, the v$_{2}$ is
passed to newly formed hadrons.

Strange quark dynamics is a useful probe of the dense matter created
at RHIC. Enhanced strangeness \cite{Rafelski} was proposed as an
important signal for the formation of the Quark-Gluon Plasma (QGP) in
nuclear collisions. The dominant production of $s\overline{s}$ pairs
via gluon-gluon interactions may lead to a strangeness (chemical and
flavor) equilibration time, comparable to the lifetime of the QGP
whereas the strangeness equilibration time in a hadronic fireball is
much longer than the fireball's lifetime. Therefore, the subsequent
hadronization of the QGP is expected to enhance the production of
strange particles. In particular, it was argued that with the
formation of QGP, the production of $\phi$ mesons is heightened.
Furthermore, $\phi$ mesons could retain information on the
condition of the hot plasma at hadronization because they
interact weakly in the hadronic matter~\cite{Shor}. The measurement of
$\phi$ mesons has been of great interest in the study of collision
dynamics and the properties of the dense matter created at
RHIC~\cite{STAR,PHENIX,DUKE-phi}. Figure~\ref{fig6} shows the elliptic
flow $v_{2}$ normalized by the valence quarks for $\phi$ mesons from
\Au\ collisions at $\sqrt{s_{NN}}$ = 200 GeV. Experimental data
\cite{PHENIX1} of $K^{+}$+$K^{-}$ and $p+\overline{p}$ are
presented for comparison.  In the intermediate $p_T$ region of
$p_T$/$n_{q}$ $>$ 0.6 GeV/$c$, the elliptic flow of charged kaons,
protons, and $\phi$ meons seems to satisfy the valence quarks scaling.
This result implies that u, d and s quarks in the initial partonic
matter formed in relativistic heavy ion collisions develop significant
collectivity with a strength characterized by $v_{2}/n_{q}$.

\FigureSeven%
\vspace*{-0.3cm}
\section{New era of heavy flavor measurements : PHENIX Silicon Vertex Tracker}
\vspace*{-0.3cm} In December 2010, the PHENIX Collaboration open new
era for measuring heavy flavor at RHIC by installing new detector
called the Silicon Vertex Tracker (VTX) \cite{RachidVTX}. The VTX is under
commissioning with beam \p\ at 500 GeV, and it will be ready to take
data this year on Run 11 at RHIC. Our main physics motivation is to
enable measurements of heavy-flavor production (charm and beauty) in
\p, \dAu\ and \Au\ collisions. Such data will illuminate the
properties of the matter created in high-energy heavy-ion
collisions. The measurements also will reveal the distribution of
gluons in protons from \p\ collisions. The VTX detector consists of
four layers of barrel detectors and covers $|\eta|~<$~1.2, and
almost a 2$\pi$ in azimuth. The inner two silicon barrels consist of
silicon pixel sensors. Figure~\ref{fig7}(a) shows ladders of a half
barrel of a pixel detector; their technology accords with that of the
ALICE1LHCB sensor-readout hybrid. The outer two barrels are silicon
stripixel detectors with a new "spiral" design, and a single-sided
sensor with 2-dimensional (X, U) readout. Figure~\ref{fig7}(b) shows
the ladders of half barrel of stripixel detector. Figure~\ref{fig7}(c)
shows VTX detector including the readout, assembled and surveyed at the
laboratory. Figure~\ref{fig7}(d) shows the VTX installed around PHENIX's
interaction point and cabled to the readout, in December 2010. 
\vspace*{-0.3cm}
\section{Summary}
\vspace*{-0.3cm} Studies of jets and anisotropic flow at RHIC have
produced very important, exciting results. These observations are open
to different interpretations and continue to test different
hypotheses, but, at the same time it seems undisputable that in \Au\
collisions at RHIC we created the deconfined and mostly
thermalized matter. The PHENIX experiment is assembling a full
picture of the results through developments in several directions:
higher-\pt, for different particle species, correlations, and full jet
reconstruction. This year, PHENIX opens new era to study the
properties of the medium, through identifying non-photonic electrons
from decays of particles carrying charm or beauty quarks, by
installing a new detector called Silicon Vertex Tracker, in December
2010. This new detector is under commissioning and will take data from RHIC
Run-11, this year.

\end{document}